\documentclass[a4paper,twoside]{article}

\usepackage{epsfig}
\usepackage{subcaption}
\usepackage{calc}
\usepackage{amssymb}
\usepackage{amstext}
\usepackage{amsmath}
\usepackage{amsthm}
\usepackage{multicol}
\usepackage{pslatex}
\usepackage{apalike}
\usepackage{algorithm2e}
\usepackage{multirow}
\usepackage{diagbox}
\usepackage{booktabs}
\usepackage{multirow}
\usepackage{placeins}
\usepackage[colorlinks=true,linkcolor=black,citecolor=black,urlcolor=black]{hyperref}

\usepackage[bottom]{footmisc}
\usepackage{SCITEPRESS}     

\usepackage{flushend}
\usepackage{microtype}

\begin{document}

\title{Temporal Modeling of Change History for Black-Box Test Suite Minimization}

\author{\authorname{Kamruzzaman Asif, Md. Siam and Kazi Sakib}
\affiliation{Institute of Information Technology, University of Dhaka, Bangladesh}
\email{\{bsse1217, bsse1104, sakib\}@iit.du.ac.bd}
}


\abstract{Test Suite Minimization (TSM) reduces the size of test suites while preserving their fault detection capability. In black-box TSM, reduction is performed without relying on production-code instrumentation. While several black-box TSM approaches have explored metrics like test logs or test similarity, these often suffer from scalability and efficiency issues. Recently, change history has been explored as a lightweight and scalable indicator for guiding black-box TSM. However, existing approaches treat historical modifications uniformly, ignoring the temporal dynamics of software evolution where recently modified code tends to be more fault-prone. To address this limitation, we introduce temporal modeling into black-box TSM and propose Temporal Risk-driven Test Suite Minimization (TRTM). TRTM extracts modification history from version-control metadata and applies exponential temporal attenuation to weight changes based on recency, producing time-weighted class-level risk scores that reflect fault-proneness. Next, it determines dependencies between test cases and production classes by constructing static call graphs derived solely from test code, preserving the black-box setting. The risk scores of the classes exercised by each test case are then aggregated using statistical measures such as Average and Geometric Mean to compute a risk score for the test case. Finally, test cases with the highest risk scores are selected to construct the reduced suite. Evaluation on a large dataset containing 14 projects with 631 versions shows that TRTM consistently outperforms the state-of-the-art baseline, achieving a mean Accuracy of 0.72 (vs.\ 0.66) and Fault Detection Rate (FDR) of 0.75 (vs.\ 0.69), while also reducing execution time.}

\onecolumn
\maketitle
\normalsize
\setcounter{footnote}{0}

\section{\uppercase{Introduction}}
\label{sec:introduction}

Modern software systems evolve continuously through frequent commits, refactorings, and feature additions. In large-scale and continuously integrated environments, executing the entire test suite for every version is often impractical due to time and resource constraints \cite{khan2018systematic}. As a result, test suite minimization (TSM) has emerged as an important technique to reduce testing effort by removing redundant test cases while preserving fault detection effectiveness \cite{yoo2012regression,khan2018systematic}. 

In recent years, black-box test suite minimization approaches have gained attention because, unlike white-box techniques, they do not require production-code analysis or structural coverage instrumentation, making them applicable to industrial systems where instrumentation is costly or unavailable \cite{philip2019fastlane,cruciani2019scalable,pan2023atm,pan2024ltm}. Existing work on black-box TSM explores diverse surrogate indicators to approximate a test case’s fault-detection potential. These include execution-history information such as commit risk and test outcome correlations \cite{philip2019fastlane}, and similarity-based diversity measures derived from test code representations \cite{cruciani2019scalable,pan2023atm,pan2024ltm}. While these approaches avoid structural instrumentation, they are limited by factors such as data availability, computational cost, or representation complexity.

More recently, software change history has been explored as a lightweight indicator for guiding black-box test suite minimization within the Change-proneness based Test suite Minimization (CTM) framework~\cite{siam2025exploratory}. However, it treats historical modifications uniformly, assigning equal importance to both recent and older changes, thereby overlooking the inherently temporal nature of software evolution. Prior studies in defect prediction and software reliability have shown that software risk exhibits temporal dynamics, where recently modified components are more likely to be fault-prone compared to components whose changes have stabilized over time \cite{graves2002predicting,nagappan2005use,hasan2005toptenlist}. Ignoring temporal recency may therefore attenuate the predictive resolution of change-based risk estimation and obscure short-term instability patterns.

To address these limitations, we introduce Temporal Risk-driven Test Suite Minimization (TRTM), an approach that integrates temporal modeling as an explicit dimension within change history-based black-box test suite minimization. TRTM first extracts change history from version-control repositories to compute class-level measures using two change metrics—change frequency and change extent—to capture how often and how substantially a class has evolved \cite{arvanitou2017method}. It then applies exponential temporal attenuation to these historical modifications, allowing their influence to decay as a function of age. This temporally weighted formulation emphasizes recent changes and produces time-weighted risk scores that reflect the likelihood of a class being fault-prone. Next, TRTM constructs test--class dependency mappings using static call-graph analysis derived solely from the test code. Finally, these time-weighted risks are aggregated using statistical operators (e.g., Average, Geometric Mean, Harmonic Mean, and Median) to compute a risk score for each test case, and the highest-scoring tests are selected to form the minimized test suite.

We systematically evaluate TRTM across a range of temporal horizons, change metrics, and aggregation operators.  
The evaluation is conducted on a large dataset containing 14 projects with 631 versions from the Defects4J benchmark. We assess fault detection effectiveness using Accuracy and Fault Detection Rate (FDR), and measure efficiency in terms of execution time. To examine the impact of temporal modeling, TRTM is compared against the state-of-the-art static change history-based 
baseline, CTM~\cite{siam2025exploratory}, under identical minimization settings.

The results show that incorporating temporal modeling consistently improves fault detection effectiveness over the static baseline. Across the explored configurations, TRTM achieves strong performance, and at the 50\% minimization budget under a representative setting, attains a mean Accuracy of 0.72 and FDR of 0.75, compared to 0.66 and 0.69 for CTM, with statistically significant improvements for both metrics. In addition, TRTM achieves lower execution time, with a mean of 0.82 minutes per version compared to 1.04 minutes for CTM. Overall, these findings demonstrate that temporal recency not only enhances the effectiveness of change history-based minimization but also delivers practical efficiency gains, making it an effective modeling dimension for black-box TSM.

\begin{figure*}[t]
  \centering
  \includegraphics[width=\textwidth]{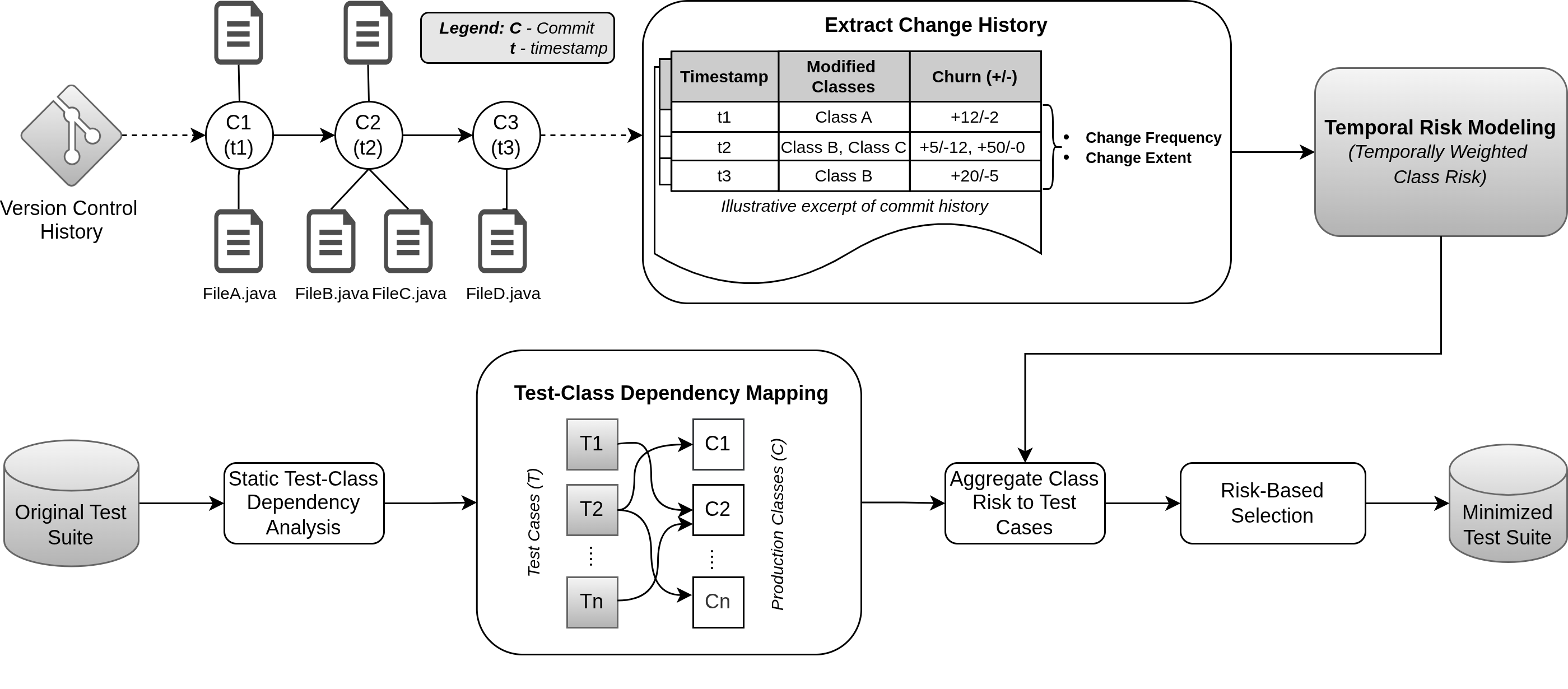}
  \caption{TRTM approach overview.}
  \label{fig:methodology}
\end{figure*}

\section{\uppercase{Related Work}}
\label{sec:related_work}

Test Suite Minimization (TSM) aims to reduce testing effort by removing redundant test cases while preserving fault detection effectiveness~\cite{yoo2012regression,khan2018systematic}. Over the years, extensive research has proposed various strategies to identify representative subsets of test suites. Broadly, these techniques fall into two categories: white-box and black-box approaches. 

White-box TSM relies on production code analysis, structural coverage information, or requirement traceability to guide test case selection. Representative techniques include greedy coverage-based heuristics~\cite{miranda2017scope,noemmer2019evaluation,mohanty2020ant}, search-based evolutionary algorithms~\cite{hemmati2013achieving,zhang2019uncertainty}, clustering methods based on coverage similarity~\cite{liu2011user,coviello2018clustering}, and model-driven or requirement-driven prioritization strategies~\cite{arrieta2019pareto}. While effective in controlled environments, such approaches require code instrumentation and coverage collection, which can introduce significant computational overhead and scalability challenges in large industrial systems~\cite{khan2018systematic}. In rapidly evolving continuous integration environments, these constraints limit the practical adoption of white-box minimization techniques.

To address these limitations, black-box TSM approaches eliminate reliance on structural coverage and instead leverage surrogate information from test code or development history. FastLane~\cite{philip2019fastlane}, the first black-box TSM technique, uses commit risk indicators, version control metadata, and historical test logs within a statistical learning framework to predict test relevance. Although FastLane achieves substantial reduction while preserving fault detection effectiveness, it depends on historical execution data, which may not always be available.

Subsequent work explored similarity-based black-box reduction. FAST-R~\cite{cruciani2019scalable} represents test cases using term-frequency vectors from test code and applies clustering to select representative tests. ATM (AST-based Test case Minimizer)~\cite{pan2023atm} further refines this approach using syntactic representations derived from abstract syntax trees and evolutionary multi-objective optimization. To improve semantic expressiveness and scalability, LTM (Language model-based Test suite Minimization)~\cite{pan2024ltm} leverages large language model embeddings to capture deeper test similarity relationships. While these approaches demonstrate strong effectiveness, they require substantial computational resources for similarity computation or embedding generation.

More recently, CTM (Change-proneness based Test suite Minimization)~\cite{siam2025exploratory} explores software change history as a lightweight and scalable source of information for black-box TSM. CTM leverages historical modifications to guide test case selection, prioritizing tests associated with frequently or substantially modified components. By relying solely on change history, it provides a practical alternative to similarity- or execution-log-based approaches.

Despite these advances, existing black-box TSM techniques—whether execution-driven or change history-based—generally treat historical information as temporally uniform. However, research in defect prediction and software reliability has consistently shown that software risk is inherently time-sensitive, with recently modified components exhibiting higher fault-proneness compared to components whose changes have stabilized over time~\cite{graves2002predicting,nagappan2005use}. 

This gap motivates the need to explicitly incorporate temporal sensitivity into change history-based black-box test suite minimization.

\section{\uppercase{Methodology}}
\label{sec:methodology}

In this study, we propose Temporal Risk-driven Test Suite Minimization (TRTM), a black-box TSM approach designed to retain a subset of test cases while preserving fault detection effectiveness. Software faults frequently arise in parts of the system that undergo structural modifications during development. TRTM therefore estimates the likelihood of each production class being fault-prone by analyzing its chronological change history from version-control metadata. These records capture both the frequency and magnitude of historical modifications. TRTM then applies exponential temporal attenuation to these modification events, weighting changes according to their recency. Consequently, classes with substantial recent modification activity receive higher risk scores, guiding which test cases are retained in the minimized suite.

As illustrated in Figure~\ref{fig:methodology}, the TRTM pipeline consists of the following stages. First, chronological modification records of production classes are extracted from version-control history. Second, class-level temporal risk is computed by applying exponential attenuation to these modification events. Third, static analysis is performed to map each test case to the production classes it exercises. Fourth, class-level risk scores of the classes exercised by each test case are aggregated using statistical operators to assign a risk score for that test case. Finally, the highest-risk test cases are selected to construct the minimized test suite. The following subsections describe each stage in detail.

\subsection{Extract Change History}
\label{subsec:change_based_class_metrics}

We begin by extracting the historical modification records of classes from the project's version-control repository. This yields a chronological view of each class's evolution, achieved by analyzing the Git commit history. For each commit, the modified classes are identified and associated change metadata is collected. Specifically, for each class $C$, modification events are recorded with: (i) the commit timestamp, and (ii) the number of lines added, deleted, or modified in that commit. These commit-level records capture two key aspects of structural evolution: how frequently a class changes (change frequency) and how extensively it evolves (change extent). Such change metrics have been widely used in prior software engineering studies as indicators of software evolution and predictors of fault-proneness~\cite{RHMANN2020419}.

Because Git records modifications at the file-path level, the same logical class may appear under different paths over time due to file renaming or relocation. To preserve class identity across such changes, we perform a consolidation step during preprocessing. We first extract change histories for all unique file paths in the repository and parse each path to determine the corresponding logical class name. When multiple file paths correspond to the same class, their records are merged by taking the union of events and ordering them chronologically by timestamp. 

After this consolidation process, each production class is represented by a unified, time-ordered sequence of modification events. These chronological histories form the basis for the temporal risk modeling described in the next subsection.

\subsection{Temporal Risk Modeling}
\label{subsec:temporal_risk_modeling}

Using the modification history extracted in the previous step, we calculate the risk (i.e., the likelihood of being fault-prone) associated with each production class by incorporating the temporal recency of its historical modifications. The intuition is that recent development activity may indicate unstable code regions and therefore deserves greater influence when calculating class-level risk.

Let a production class $C$ have a sequence of $n$ historical modification events. Each event $i$ occurred at time $t_i$. For a target version evaluated at time $t$, we define the age of modification event $i$ as

\begin{equation}
\Delta t_i = t - t_i ,
\label{eq:deltat}
\end{equation}

where $\Delta t_i$ represents the elapsed time since the modification occurred, measured in days.

To account for temporal recency, the contribution of each modification event is attenuated as a function of its age. Specifically, we estimate the temporal risk of class $C$ as

\begin{equation}
R_{\alpha}(C) = \sum_{i=1}^{n} w_i \, e^{-\alpha \Delta t_i},
\label{eq:r_alpha}
\end{equation}

where $w_i$ denotes the magnitude of modification event $i$, and $\alpha > 0$ is a temporal attenuation parameter controlling how rapidly the influence of older modifications decays over time. This formulation assigns greater weight to recent modifications while gradually reducing the impact of older events. Similar exponential attenuation mechanisms have been employed in prior defect prediction and reliability studies to model the diminishing influence of historical development activity~\cite{graves2002predicting,hasan2005toptenlist}.

The event weight $w_i$ depends on the change metric used to characterize the modification. In this study, we consider two complementary change metrics commonly used in software evolution analysis.

\textbf{Change Frequency.}
In this formulation, each modification event contributes equally 
to the risk estimation:
\begin{equation}
w_i = 1.
\label{eq:w_i_freq}
\end{equation}
This yields the temporally weighted change frequency:
\begin{equation}
R_{\alpha}^{\mathrm{Freq}}(C) = \sum_{i=1}^{n} e^{-\alpha \Delta t_i}.
\label{eq:r_alpha_freq}
\end{equation}

\textbf{Change Extent.}
Here, the contribution of each modification event is weighted by the magnitude of code churn introduced in that commit. We define the churn of modification event $i$ as
\begin{equation}
\mathrm{Churn}_i = \mathrm{Add}_i + \mathrm{Del}_i + \mathrm{Mod}_i,
\label{eq:churn}
\end{equation}
where $\mathrm{Add}_i$, $\mathrm{Del}_i$, and $\mathrm{Mod}_i$ denote 
lines added, deleted, and modified in event $i$, respectively. 
To mitigate the heavy-tailed distribution typically observed in such 
metrics, we apply logarithmic normalization:
\begin{equation}
w_i = \ln \left(1 + \mathrm{Churn}_i \right).
\label{eq:w_i_ext}
\end{equation}
The resulting temporally weighted change extent is given by
\begin{equation}
R_{\alpha}^{\mathrm{Ext}}(C) =
\sum_{i=1}^{n}
\ln \left(1 + \mathrm{Churn}_i \right)
e^{-\alpha \Delta t_i}.
\label{eq:r_alpha_ext}
\end{equation}

The attenuation parameter $\alpha$ determines how strongly the model prioritizes recent modifications over older ones. Rather than fixing a single value, we systematically evaluate a range of decay configurations spanning multiple temporal scales to analyze the impact of temporal sensitivity on minimization effectiveness. For interpretability, $\alpha$ is parameterized using a half-life formulation

\begin{equation}
\alpha = \frac{\ln(2)}{T},
\label{eq:alpha_halflife}
\end{equation}

where $T$ denotes the temporal half-life of a modification's risk. Under this formulation, the influence of a modification decreases to 50\% of its original value after $T$ days. Varying $T$ allows the model to systematically examine how different temporal sensitivities influence risk estimation, from short-term recency to long-term historical effects.

\begin{figure}[htbp]
  \centering
  \includegraphics[width=0.9\columnwidth]{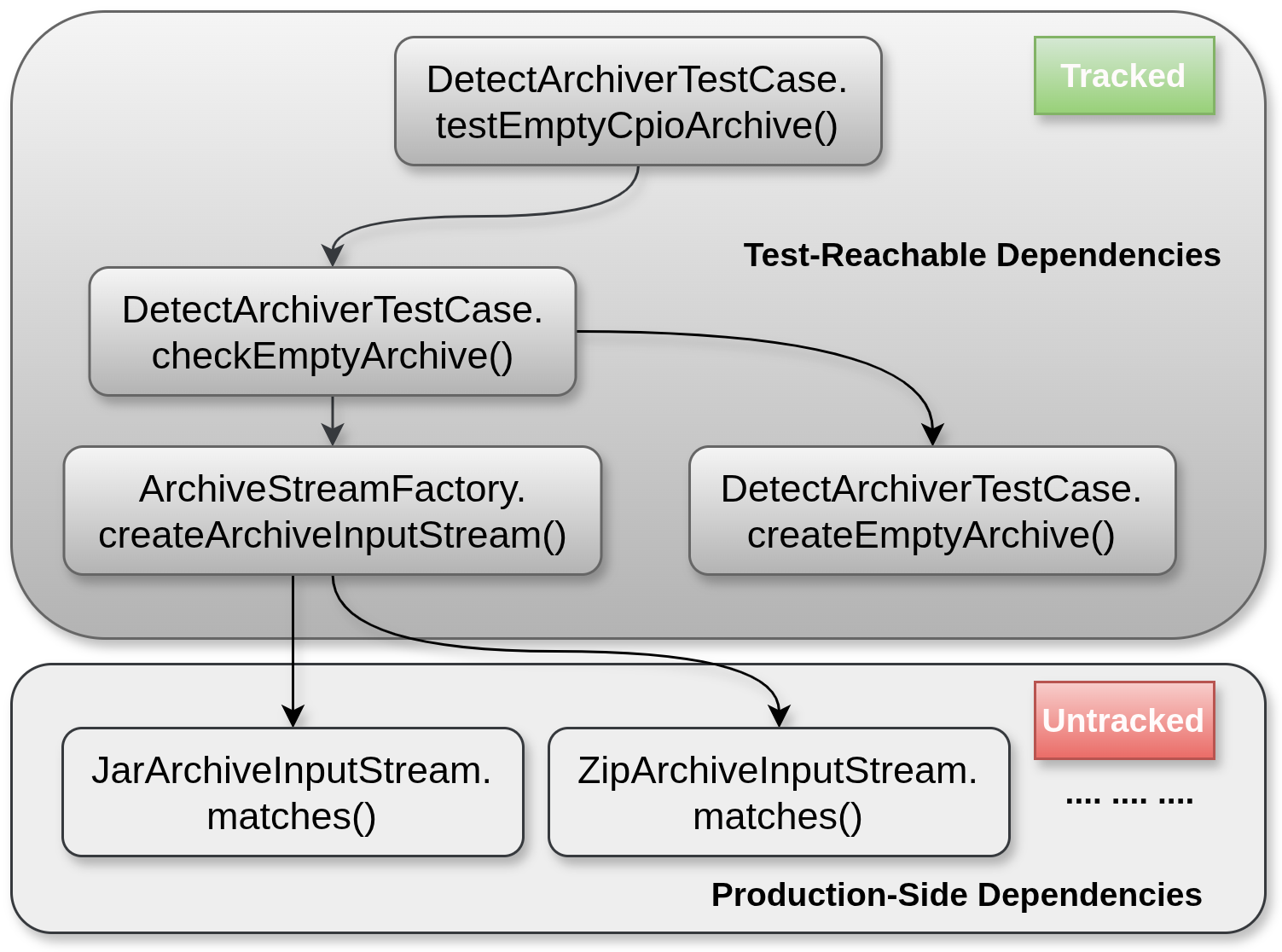}
  \caption{Test-Class Dependencies.}
  \label{fig:test-class_dependencies}
\end{figure}

\subsection{Extract Test--Class Dependencies}
\label{subsec:static_dependency_extraction}

To associate class-level temporal risk with test cases, we identify which production classes are exercised by each test case. This step establishes the structural relationships required to associate test cases with the classes whose risk they potentially expose. Consistent with the black-box setting, we derive this mapping using static analysis of the test code without instrumenting or modifying the production system.

For our analysis, we generate call graphs (gray-marked portion of Figure \ref{fig:test-class_dependencies}) strictly from the test suite of each project using the \texttt{java-callgraph} tool. The test code is first compiled into JAR files, which serve as a direct input to construct the call graphs. We opt for static call graphs because they are computationally efficient and highly suitable for a black-box setup. Unlike dynamic call graphs, which require executing the program and monitoring runtime behavior, static call graphs can be derived directly from compiled bytecode. This avoids the need for deep source code analysis or system instrumentation, aligning with our goal of minimizing test suites in a production-agnostic manner.

Treating each test method as an entry point in the call graph, we perform a Depth-First Search (DFS) to traverse the invocation edges and collect all reachable associated classes. This DFS traversal ensures we systematically capture both direct invocations and deep transitive dependencies between the test suite and the production classes.

Formally, for a test case $T$, we define its static dependency set as:

\begin{equation}
\mathrm{Deps}(T) = \{C_1, C_2, \dots, C_m\},
\label{eq:deps}
\end{equation}

where each $C_j$ denotes a production class that is directly or indirectly reachable from $T$ via the static call graph.
The resulting test--class dependency mappings provide the structural foundation required to aggregate our temporally weighted class risks to individual test cases in the subsequent step.

\subsection{Aggregate Class Risk to Test Cases}
\label{subsec:aggregation_to_test_level}

The dependency mappings defined in Equation~\ref{eq:deps} identify the production classes exercised by each test case. Using these mappings, class-level temporal risk scores are aggregated for each test case.

Let $T$ denote a test case and $\mathrm{Deps}(T)=\{C_1,C_2,\dots,C_m\}$ denote the set of production classes reachable from $T$ through the static call graph. The score of test case $T$ is computed by aggregating the temporal 
risk scores of all classes in $\mathrm{Deps}(T)$:

\begin{equation}
\mathrm{Score}(T)=A(\{R_{\alpha}(C)\mid C\in\mathrm{Deps}(T)\}),
\label{eq:test_score}
\end{equation}

where $R_{\alpha}(C)$ denotes the temporally weighted risk score of class $C$ computed in Section~\ref{subsec:temporal_risk_modeling}, and $A(\cdot)$ represents an aggregation operator applied to the set of class-level risk scores.

Several statistical operators can be used to summarize class-level risk distributions, including Minimum, Maximum, Sum, Average, Geometric Mean, Harmonic Mean, Median, and Standard Deviation. However, operators such as Minimum and Maximum are highly sensitive to outliers and may assign disproportionate influence to extreme class risks \cite{ahmadi2025robust,rousseeuw2011robust}. Similarly, Standard Deviation primarily captures variability rather than central tendency and can be unstable for skewed distributions \cite{rousseeuw2011robust}. The Sum operator may also bias the score toward test cases with a large number of dependencies, potentially overshadowing smaller dependency sets that contain highly fault-prone classes.

To obtain stable and robust test-level risk estimates, four aggregation operators are considered: Arithmetic Mean (Avg), Geometric Mean (GMean), Harmonic Mean (HMean), and Median. These operators provide complementary perspectives on central tendency while remaining less sensitive to extreme values in the class-level risk distribution.

\subsection{Risk-Based Selection}
\label{subsec:risk_based_selection}

The aggregation process described in Section~\ref{subsec:aggregation_to_test_level} produces a test-level risk score for each test case. These scores are used to guide budget-constrained test suite minimization. For each project, we compute the test-level risk scores for all test cases. We then select the subset of test cases with the highest score values under a predefined retention budget. Consistent with prior black-box TSM studies, we evaluate three commonly adopted budget levels---25\%, 50\%, and 75\% of the original test suite~\cite{pan2023atm,pan2024ltm,siam2025exploratory}. The selected subset constitutes the minimized test suite.

\section{\uppercase{Experimental Setup and Result Analysis}}
\label{sec:experiment_results}

In this section, we evaluate TRTM through a series of empirical experiments. We begin by outlining the research questions that guide our study, followed by a description of the experimental design, including the subject systems, configurations, and evaluation metrics. We then present and analyze the results obtained from applying TRTM to the test suites.

\subsection{Research Questions}
\label{subsec:research_questions}

To systematically examine the role of temporal modeling in test suite minimization, our study is guided by the following research questions:

\noindent\textbf{RQ1:} \emph{How does temporal modeling influence the effectiveness of test suite minimization under different configurations?}

This research question investigates the sensitivity of TRTM to its configuration parameters, examining how variations in temporal decay horizons, change metrics, and aggregation operators influence fault detection effectiveness. Since different configurations may lead to different outcomes, we evaluate TRTM across these dimensions to understand how temporal sensitivity and risk aggregation affect the preservation of fault-revealing test cases and to identify stable and effective configuration settings.

\noindent\textbf{RQ2:} \emph{Does temporal modeling improve the effectiveness and efficiency of test suite minimization compared with the state-of-the-art static approach?}

This research question evaluates whether incorporating temporal modeling improves fault detection effectiveness and computational efficiency. To answer this, we take the representative TRTM configuration 
identified in RQ1 and compare it directly against the 
state-of-the-art baseline, CTM~\cite{siam2025exploratory}, 
under identical budgets and experimental settings.

\subsection{Experimental Design}
\label{subsec:experimental_design}

To evaluate TRTM, we conduct a systematic comparison against CTM, the state-of-the-art static change history-based approach, across multiple configurations. All experiments were executed on a machine equipped with an Intel Core i7-9750H processor (6 cores, 12 threads; base frequency 2.60~GHz with turbo up to 4.50~GHz), 30~GB RAM, running Ubuntu 24.04.4 LTS. The implementation is deterministic, yielding identical results across runs.

\noindent\textit{1) \textbf{Configurations}:} A configuration is defined by a combination of change metric, temporal horizon, and aggregation operator used to compute test case risk scores. We consider two class-level change metrics: \textit{Change Frequency} and \textit{Change Extent}. These metrics are extended using temporal attenuation as described in Section~\ref{subsec:temporal_risk_modeling}. 

To systematically analyze the influence of temporal recency, we evaluate the temporal attenuation parameter across a base-2 exponential progression of temporal horizons ranging from $2^0$ to $2^9$ days (1--512 days). Such logarithmic parameter sweeps are commonly used to explore sensitivity across multiple orders of magnitude while avoiding bias toward specific ranges~\cite{hutter2011sequential,bergstra2012random}. The upper bound of 512 days deliberately exceeds the empirically validated predictive windows reported in defect prediction literature~\cite{graves2002predicting}, ensuring that the full relevant temporal range is covered without imposing arbitrary horizon choices. For each test case, class-level risk scores of the exercised classes are aggregated using the statistical operators defined in Section~\ref{subsec:aggregation_to_test_level}. In total, this results in $2 \times 10 \times 4 = 80$ TRTM configurations (2 change metrics, 10 temporal horizons, and 4 aggregation operators) evaluated across all project versions.

\noindent\textit{2) \textbf{Minimization Budgets}:}
The minimization budget defines the proportion of test cases retained after reduction relative to the original test suite. Consistent with prior black-box TSM studies, we evaluate three retention levels: 25\%, 50\%, and 75\% \cite{pan2023atm}. For each configuration and budget level, test cases with the highest association scores are selected until the specified retention percentage is reached.

\subsection{Dataset}
\label{subsec:dataset}

We evaluate TRTM using Defects4J\footnote{\texttt{https://github.com/rjust/defects4j}} (v2.0.1), a widely adopted benchmark in software testing research~\cite{rjust2014defects4j}. Defects4J provides real, reproducible faults from open-source Java projects along with their corresponding test suites and fault-fixing commits. Each buggy version contains a single real fault that triggers failures in one or more test cases.

The initial dataset includes 16 projects with 661 buggy versions. We exclude \textit{Chart} due to unreliable version-control history that prevented reliable extraction of historical change information, and \textit{Collections} due to version inconsistencies and its limited size of only 4 buggy versions, which limits meaningful empirical evaluation. The final dataset consists of 14 projects and 631 buggy versions. For each buggy version, change history is extracted only from preceding commits to avoid using future information.

Table~\ref{tab:dataset} summarizes the characteristics of the subject systems. Project sizes range from 2--74~KLoC, with test suites spanning 4--73~KLoC. The number of buggy versions per project varies between 6 and 112, and test suites contain between 152 and 3,918 test cases per version on average. Statistics are derived from repository metadata and CLOC analysis. By using the same benchmark adopted in prior black-box TSM studies \cite{pan2023atm,pan2024ltm,siam2025exploratory}, the evaluation ensures compatibility with existing work while covering diverse system sizes and evolution histories.

\begin{table}[t]
\centering
\caption{Subject projects statistics (Defects4J v2.0.1).}
\label{tab:dataset}
\setlength{\tabcolsep}{4pt}
\resizebox{\columnwidth}{!}{%
\begin{tabular}{lccccc}
\toprule
Project & Size & \# Versions & Test Size & Avg \# Tests & Avg \# Commits \\
        & (KLoC) & (faults) & (KLoC) & per version & per version \\
\midrule
Cli             & 2  & 39  & 4  & 256  & 486 \\
Codec           & 9  & 18  & 15 & 413  & 928 \\
Compress        & 45 & 47  & 29 & 404  & 1,393 \\
Csv             & 2  & 16  & 7  & 193  & 828 \\
Gson            & 9  & 18  & 20 & 984  & 1,226 \\
JacksonCore     & 31 & 26  & 45 & 356  & 902 \\
JacksonDatabind & 74 & 112 & 72 & 1,814 & 3,009 \\
JacksonXml      & 6  & 6   & 10 & 152  & 660 \\
Jsoup           & 14 & 93  & 13 & 494  & 785 \\
JxPath          & 20 & 22  & 6  & 250  & 374 \\
Lang            & 30 & 64  & 61 & 1,796 & 2,398 \\
Math            & 71 & 106 & 73 & 2,078 & 2,932 \\
Mockito         & 21 & 38  & 36 & 1,182 & 1,880 \\
Time            & 30 & 26  & 56 & 3,918 & 1,600 \\
\bottomrule
\end{tabular}%
}
\end{table}

\begin{figure*}[t]
\centering
\includegraphics[width=\textwidth]
  {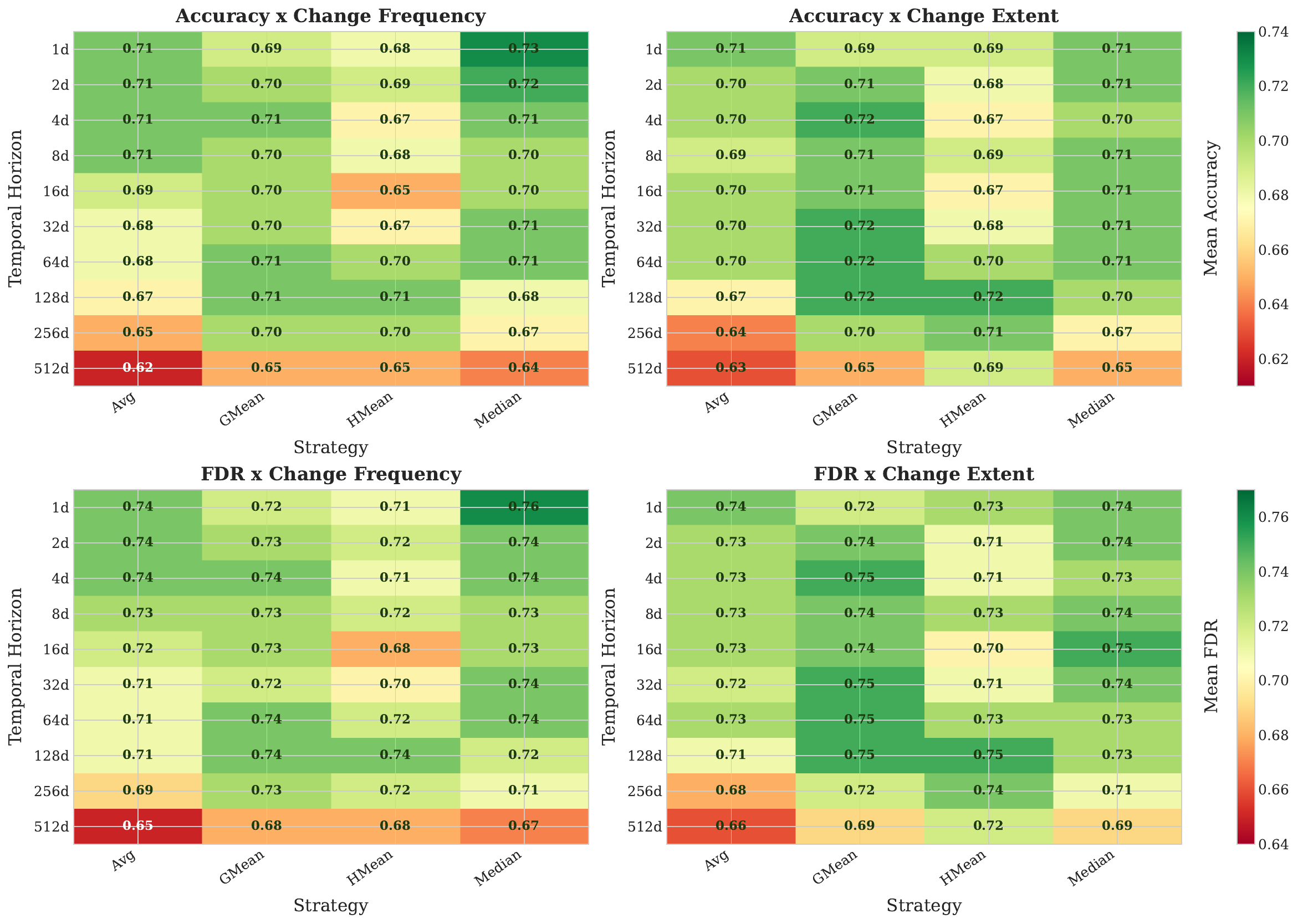}
\caption{Four-panel heatmap of TRTM performance across 
temporal horizons at the 50\% minimization budget. 
Rows correspond to temporal horizons (1--512 days at 
base-2 intervals); columns correspond to aggregation 
operators (\textit{Avg}, \textit{GMean}, \textit{HMean}, 
\textit{Median}). Cell values report mean Accuracy 
(top panels) and FDR (bottom panels) across all 
project versions for \textit{Change Frequency} (left) 
and \textit{Change Extent} (right) metrics.}
\label{fig:temporal_horizon_4panel_budget50}
\end{figure*}

\flushbottom
\subsection{Evaluation Metrics and Statistical Analysis}
\label{subsec:evaluation_metrics}

To evaluate the performance of TRTM, we consider three metrics: Accuracy, Fault Detection Rate (FDR), and Execution Time.

\noindent\textit{1) \textbf{Accuracy}:}
The primary objective of test suite minimization is to reduce the number of test cases while maintaining the fault detection capability. Accuracy measures the proportion of fault-revealing test cases retained in the minimized test suite relative to the original suite. Formally, accuracy is defined as:
\begin{equation}
\mathrm{Accuracy} = \frac{|F'|}{|F|},
\label{eq:accuracy}
\end{equation}
where $|F'|$ denotes the number of fault-revealing test cases retained in the minimized test suite, and $|F|$ represents the total number of fault-revealing test cases in the original test suite for a given buggy version. Accuracy therefore quantifies how well the minimized suite preserves the defect-exposing capacity of the original test suite at the test-case level.

\noindent\textit{2) \textbf{Fault Detection Rate (FDR)}:}
While accuracy reflects proportional preservation, Fault Detection Rate (FDR) provides a more relaxed perspective by evaluating whether at least one fault-revealing test case is retained after minimization. FDR measures the percentage of buggy versions for which the minimized test suite contains at least one fault-revealing test case \cite{pan2023atm}. It is defined as:
\begin{equation}
\mathrm{FDR} = \frac{1}{n} \sum_{i=1}^{n} f_i,
\label{eq:fdr}
\end{equation}
where $n$ denotes the total number of buggy versions and
\begin{equation}
f_i =
\begin{cases}
1, & \parbox[t]{0.72\linewidth}{\raggedright if the minimized suite for version $i$ retains at least one fault-revealing test case,} \\
0, & \text{otherwise.}
\end{cases}
\label{eq:fi}
\end{equation}

\noindent\textit{3) \textbf{Execution Time}:}
To assess the efficiency of TRTM, we measure the total execution time, which includes the time required for change-history extraction, call-graph generation, test--class dependency analysis, and risk scoring.

\noindent \textbf{Statistical Analysis:}
To assess whether TRTM leads to statistically significant improvements over the baseline approach, we perform paired comparisons at the project-version level. For continuous effectiveness measures, we apply the Wilcoxon signed-rank test, a non-parametric test suitable for paired samples. For the binary fault detection outcome, we use Fisher’s exact test to compare the proportion of detected faults. Statistical tests are conducted between TRTM and the baseline under identical minimization budgets (25\%, 50\%, and 75\%). To control for multiple comparisons, Bonferroni correction is applied to adjust the significance threshold. Cliff’s delta ($\delta$) is reported to provide an estimate of effect size.

\subsection{Result Analysis}
\label{subsec:results_summary}

Analysis focuses on results obtained at the 50\% 
minimization budget, as trends observed at the 25\% and 
75\% budgets are consistent and are reported in the 
replication package \cite{kamruzzaman_asif_2026_19506050}. The first part of the analysis examines how effectiveness varies across different temporal decay horizons, change metrics, and aggregation operators, capturing the sensitivity of TRTM to its configuration choices (RQ1). The second part compares the best-performing TRTM configuration against the state-of-the-art static baseline approach~(RQ2).

\subsubsection{RQ1: Influence of Temporal Modeling}

Figure~\ref{fig:temporal_horizon_4panel_budget50} 
summarizes the performance of TRTM across temporal 
decay horizons, aggregation operators, and change 
metrics at the 50\% minimization budget. The figure 
provides an overview of how minimization effectiveness 
varies as temporal sensitivity changes across the 
evaluated configurations.

A consistent trend is visible across both change 
metrics and all aggregation operators. Configurations 
with shorter temporal horizons generally achieve 
higher effectiveness, while performance gradually 
declines as the horizon increases. In particular, 
results remain strongest within the lower portion of 
the evaluated range (approximately 1--128 days) and 
decrease toward the longest horizons. For example, 
under the \textit{GMean}/\textit{ChangeExtent} 
configuration, mean Accuracy reaches approximately 
0.72 at intermediate horizons but drops to 0.65 at 
the 512-day horizon. This pattern indicates that 
temporal modeling is most beneficial when recent 
modifications receive stronger emphasis, whereas very 
long horizons increasingly resemble static change 
aggregation.

Although the horizon effect is visible across all operators, their behaviors differ. The \textit{Avg} operator shows a steady performance decline as the horizon increases. Under \textit{ChangeExtent}, mean Accuracy decreases from approximately 0.71 at the 1-day horizon to around 0.63 at the 512-day horizon. This pattern reflects the arithmetic mean’s tendency to dilute the influence of recent modifications as older changes accumulate. In contrast, the \textit{HMean} operator exhibits unstable behavior with irregular drops at specific horizons. For example, under \textit{ChangeFrequency}, mean Accuracy falls to about 0.65 at the 16-day horizon while adjacent horizons remain near 0.67--0.68. This instability occurs because the harmonic mean is highly sensitive to near-zero values; when temporal decay produces sparse risk distributions, even a single low-risk class can substantially suppress the aggregated score.

The remaining operators, \textit{GMean} and 
\textit{Median}, exhibit more consistent behavior. 
\textit{Median} achieves the highest peak Accuracy 
(0.73) at the 1-day horizon under 
\textit{ChangeFrequency}, but this performance 
declines steadily as the horizon increases. The strong performance at very short horizons likely reflects the close temporal proximity 
between code modifications and fault introduction in the Defects4J dataset, where buggy versions often occur shortly after the changes that introduced the fault. In contrast, \textit{GMean} maintains stable Accuracy between approximately 0.71 and 0.72 across a wide 
range of horizons (2--128 days), indicating a more 
stable aggregation behavior.

Table~\ref{tab:gmean_vs_median_stats} further depicts 
the stability of these two operators across all project 
versions. Although \textit{Median} achieves slightly higher 
mean values (Accuracy 0.73 and FDR 0.76), its minimum 
Accuracy drops to 0.48, indicating larger variability 
across projects. In contrast, \textit{GMean} maintains a 
higher minimum Accuracy of 0.53 while achieving comparable 
mean performance (Accuracy 0.72 and FDR 0.75), suggesting 
a more stable performance floor. This variability can also 
be observed in individual projects. For example, in 
JacksonCore, \textit{Median}/\textit{ChangeFrequency} 
(1-day horizon) yields an Accuracy of 0.48, whereas 
\textit{GMean}/\textit{ChangeExtent} (32-day horizon) 
achieves a higher Accuracy of 0.69. Such differences arise because 
\textit{Median} is sensitive to the underlying distribution of 
values, while \textit{GMean} captures multiplicative interactions 
across contributing factors, often resulting in more balanced 
aggregation behavior~\cite{huber2009robust,fleming1986geometric}. 
Detailed per-project results are available in the 
replication package \cite{kamruzzaman_asif_2026_19506050}. Based on these observations, \textit{GMean} is selected as the aggregation operator for subsequent analysis.

\begin{table}[htbp]
\centering
\caption{Descriptive statistics for 
\textit{GMean}/\textit{ChangeExtent} (32-day horizon) 
and \textit{Median}/\textit{ChangeFrequency} (1-day 
horizon) across all project versions at the 50\% 
minimization budget.}
\label{tab:gmean_vs_median_stats}
\resizebox{\columnwidth}{!}{%
\small
\begin{tabular}{lcccc}
\toprule
& \multicolumn{2}{c}{\textbf{Accuracy}} 
& \multicolumn{2}{c}{\textbf{FDR}} \\
\cmidrule(lr){2-3}\cmidrule(lr){4-5}
\textbf{Statistics} 
& \textit{GMean} & \textit{Median} 
& \textit{GMean} & \textit{Median} \\
\midrule
Min        & \textbf{0.53} & 0.48 
           & \textbf{0.59} & 0.52 \\
25\% (Q1)  & \textbf{0.64} & \textbf{0.64} 
           & \textbf{0.65} & \textbf{0.67} \\
Mean       & 0.72          & \textbf{0.73} 
           & 0.75          & \textbf{0.76} \\
Median     & \textbf{0.73} & 0.71 
           & \textbf{0.75} & \textbf{0.75} \\
75\% (Q3)  & 0.77          & \textbf{0.81} 
           & 0.82          & \textbf{0.85} \\
Max        & 1.00          & 1.00          
           & 1.00          & 1.00 \\
\bottomrule
\end{tabular}%
}
\end{table}

\begin{figure}[t]
\centering
\includegraphics[width=0.95\linewidth]
  {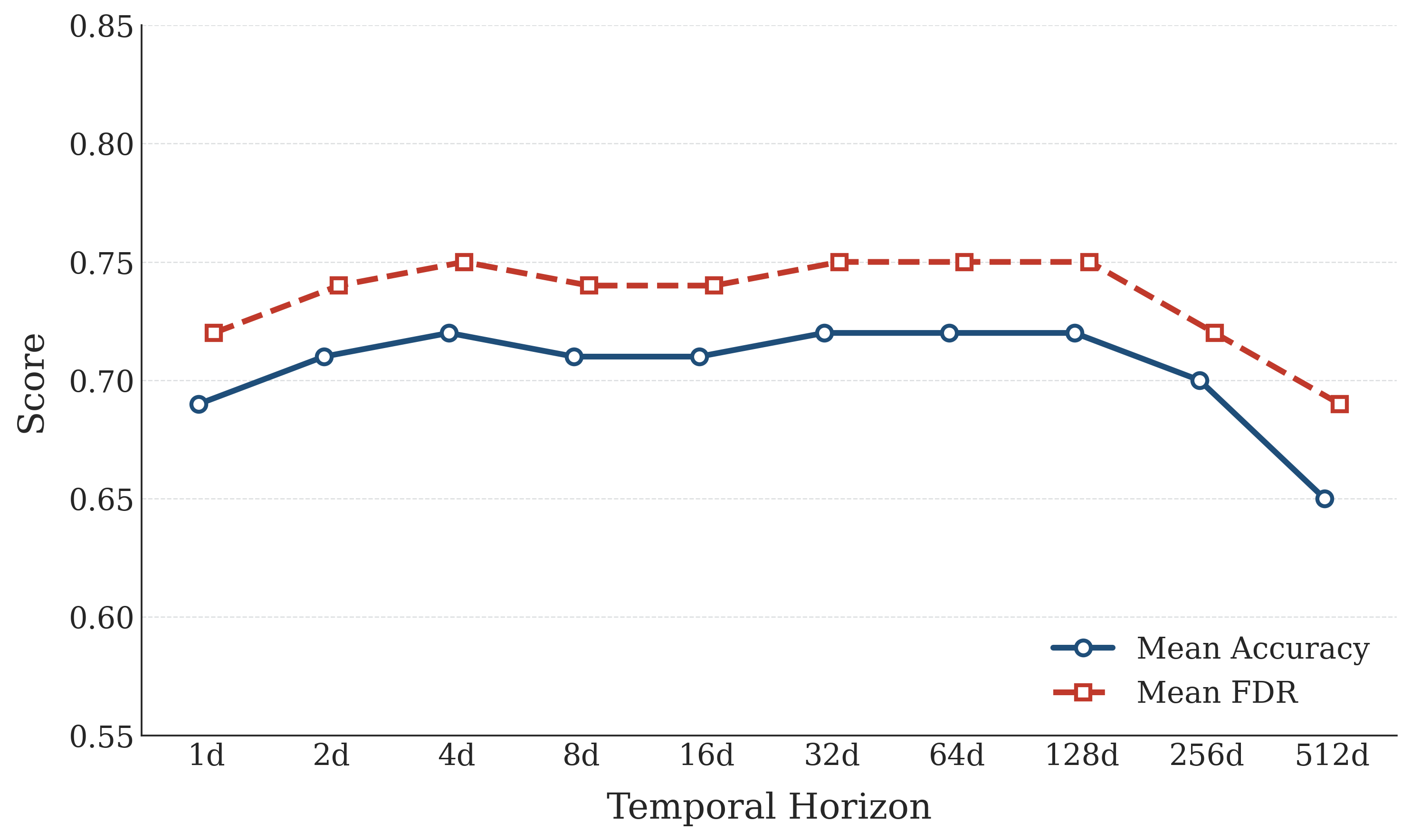}
\caption{Mean Accuracy and FDR of TRTM 
(\textit{GMean}/\textit{Change Extent}) as a function 
of temporal horizon at the 50\% minimization budget.}
\label{fig:temporal_horizon_gmean_budget50}
\end{figure}

Across the two change metrics, \textit{ChangeExtent} 
consistently produces slightly stronger and more stable 
results when combined with \textit{GMean}. Figure~
\ref{fig:temporal_horizon_gmean_budget50} shows how 
performance varies with the temporal horizon for this 
configuration. Both Accuracy and FDR increase slightly 
from very short horizons, remain stable across an 
intermediate range (approximately 4--128 days), and 
decline beyond this region. Selecting a configuration from within 
such a stable performance plateau is a principled 
approach in empirical parameter analysis, as 
configurations within a flat region yield equivalent 
and reliable performance~\cite{bergstra2012random,tantithamthavorn2016automated}. Based on this behavior, a 
\textbf{32-day horizon} is selected for the subsequent 
analysis, as it lies within the stable performance 
plateau while representing a practically meaningful 
development window that captures recent modification 
activity across the subject systems.

\vspace{0.5\baselineskip}
\noindent\fbox{\parbox{0.96\linewidth}{
\textbf{Answer to RQ1:} Incorporating temporal modeling 
into test suite minimization meaningfully 
influences fault detection effectiveness, with the 
degree of impact shaped by both the decay horizon and 
the choice of aggregation operator. Across configurations, shorter temporal horizons consistently achieve higher fault-detection effectiveness, while very long horizons degrade performance as they increasingly resemble static change aggregation. Among the evaluated aggregation operators, \textit{GMean} provides the most stable behavior across projects, and \textit{ChangeExtent} yields slightly stronger results than \textit{ChangeFrequency} (achieving a mean Accuracy of 0.72 and an FDR of 0.75 at the 32-day horizon).
}}

\subsubsection{RQ2: Comparison with Baseline Approach}

We compare the effectiveness and efficiency of TRTM with the current state-of-the-art static baseline approach CTM, using the best TRTM configuration identified in RQ1.

\noindent\textbf{Effectiveness.}
Table~\ref{tab:effectiveness_50_budget} presents the 
per-project Accuracy and FDR results at the 50\% 
minimization budget. TRTM consistently outperforms CTM 
across the project set, achieving a mean Accuracy of 
0.72 and mean FDR of 0.75, compared to 0.66 and 0.69 
for CTM. The improvement is broad-based: TRTM achieves 
higher or equal Accuracy in 12 of 14 projects and 
higher or equal FDR in 13 of 14 projects. The minimum 
Accuracy also improves from 0.42 to 0.53, indicating 
that temporal modeling reduces worst-case minimization 
failures across the project set.

Several projects exhibit particularly notable 
improvements, revealing a consistent pattern in how 
temporal modeling enhances effectiveness. For example, 
\textit{JacksonXml} achieves perfect Accuracy and FDR of 
1.00, compared to 0.80 for CTM. Similarly, \textit{Jsoup} 
and \textit{Csv} show substantial gains over CTM (e.g., 
\textit{Jsoup}: 0.42→0.53 Accuracy and 0.48→0.60 FDR; 
\textit{Csv}: 0.53→0.69 Accuracy). Similar improvements are observed in \textit{Mockito}, where FDR increases from 0.79 to 0.86. These improvements indicate that temporal modeling more accurately captures the relationship between recent code changes and fault-proneness, enabling more effective identification of fault-relevant test cases compared to the static baseline. TRTM yields marginally lower Accuracy than CTM only for \textit{Cli} (0.77 vs.\ 0.78) and \textit{JacksonCore} (0.67 vs.\ 0.71), both of which reflect the limitation of a fixed global horizon — for instance, our extended analysis shows that \textit{Cli} achieves Accuracy of 0.80 at shorter horizons of 2 and 4 days, suggesting that per-project horizon calibration could further improve performance.

\vspace{0.5\baselineskip}
\begin{table}[htbp]
\centering
\caption{Accuracy, FDR, and Execution Time of TRTM (GMean/Change Extent, 
32-day horizon) and CTM (GMean/Change Extent) 
across projects at the 50\% minimization budget. Bold 
indicates the higher value per metric per project (for Time, lower is better).}
\label{tab:effectiveness_50_budget}
\resizebox{\columnwidth}{!}{%
\Huge
\begin{tabular}{l c c c c c c}
\toprule
\multirow{2}{*}{\textbf{Project}} & 
\multicolumn{2}{c}{\textbf{Accuracy}} & 
\multicolumn{2}{c}{\textbf{FDR}} &
\multicolumn{2}{c}{\textbf{Time (min)}} \\
\cmidrule(lr){2-3} \cmidrule(lr){4-5} \cmidrule(lr){6-7}
& \textbf{TRTM} & \textbf{CTM} & 
  \textbf{TRTM} & \textbf{CTM} &
  \textbf{TRTM} & \textbf{CTM} \\
\midrule
Time            & \textbf{0.73} & 0.66          & \textbf{0.75} & 0.71          & \textbf{1.19} & 1.35 \\
JxPath          & \textbf{0.55} & \textbf{0.55} & \textbf{0.59} & \textbf{0.59} & 0.35          & \textbf{0.34} \\
Codec           & \textbf{0.77} & 0.72          & \textbf{0.81} & 0.75          & \textbf{0.55} & 0.67 \\
JacksonXml      & \textbf{1.00} & 0.80          & \textbf{1.00} & 0.80          & \textbf{0.27} & 0.44 \\
JacksonDatabind & \textbf{0.84} & \textbf{0.84} & \textbf{0.86} & 0.85          & \textbf{1.68} & 2.06 \\
Gson            & \textbf{0.75} & 0.68          & \textbf{0.75} & \textbf{0.75} & \textbf{0.73} & 0.93 \\
Math            & \textbf{0.63} & 0.53          & \textbf{0.65} & 0.57          & \textbf{1.83} & 2.46 \\
Jsoup           & \textbf{0.53} & 0.42          & \textbf{0.60} & 0.48          & \textbf{0.44} & 0.55 \\
Compress        & \textbf{0.78} & 0.75          & \textbf{0.79} & 0.77          & \textbf{0.91} & 1.05 \\
Mockito         & \textbf{0.74} & 0.68          & \textbf{0.86} & 0.79          & \textbf{1.18} & 1.49 \\
Cli             & 0.77          & \textbf{0.78} & \textbf{0.83} & \textbf{0.83} & \textbf{0.22} & 0.35 \\
Csv             & \textbf{0.69} & 0.53          & \textbf{0.69} & 0.56          & \textbf{0.43} & 0.58 \\
JacksonCore     & 0.67          & \textbf{0.71} & 0.67          & \textbf{0.71} & \textbf{0.45} & 0.61 \\
Lang            & \textbf{0.61} & 0.53          & \textbf{0.63} & 0.57          & \textbf{1.20} & 1.64 \\
\midrule
\multicolumn{7}{l}{\textbf{Statistics}} \\
\midrule
Min             & \textbf{0.53} & 0.42          & \textbf{0.59} & 0.48          & \textbf{0.22} & 0.34 \\
25\% Quartile   & \textbf{0.64} & 0.54          & \textbf{0.65} & 0.58          & \textbf{0.43} & 0.56 \\
Mean            & \textbf{0.72} & 0.66          & \textbf{0.75} & 0.69          & \textbf{0.82} & 1.04 \\
Median          & \textbf{0.73} & 0.68          & \textbf{0.75} & 0.73          & \textbf{0.64} & 0.80 \\
75\% Quartile   & \textbf{0.77} & 0.74          & \textbf{0.82} & 0.78          & \textbf{1.19} & 1.45 \\
Max             & \textbf{1.00} & 0.84          & \textbf{1.00} & 0.85          & \textbf{1.83} & 2.46 \\
\bottomrule
\end{tabular}%
}
\end{table}
\vspace{0.3\baselineskip}

Statistical analysis at the 50\% minimization budget 
confirms that TRTM yields significant improvements 
over CTM on both effectiveness metrics. Fisher's exact 
test shows a statistically significant improvement in 
FDR ($p < 0.001$, OR~$= 7.27$), indicating that TRTM 
is substantially more likely to retain at least one 
fault-detecting test case in versions where CTM fails. 
The Wilcoxon signed-rank test further confirms a 
statistically significant improvement in Accuracy after 
Bonferroni correction ($p = 0.004$, Cliff's $\delta = 0.063$). 
While the effect size for Accuracy is small, the improvement 
is consistent across 631 project versions, indicating a systematic benefit of temporal modeling. Results at 25\% and 75\% budgets follow consistent trends and are available in the replication package.

\noindent\textbf{Efficiency.}
TRTM achieves efficient execution while improving effectiveness 
over CTM. The mean execution time per version is 0.82 minutes 
for TRTM compared to 1.04 minutes for CTM (median: 0.64 vs.\ 
0.80 minutes), indicating an overall reduction in execution 
time across projects. These results indicate that incorporating 
temporal attenuation—a lightweight weighting operation—preserves efficient execution while delivering statistically significant gains in test suite minimization effectiveness.

\vspace{0.5\baselineskip}
\noindent\fbox{\parbox{0.96\linewidth}{
\textbf{Answer to RQ2:} TRTM consistently improves fault 
detection effectiveness over the state-of-the-art static CTM baseline. At 
the 50\% minimization budget, TRTM achieves a mean Accuracy 
of 0.72 (vs.\ 0.66) and an FDR of 0.75 (vs.\ 0.69), while 
elevating the minimum Accuracy floor to 0.53 (vs.\ 0.42) 
across the project set. These improvements are statistically 
significant for both metrics after Bonferroni correction 
(Accuracy: $p = 0.004$; FDR: $p < 0.001$, OR~$= 7.27$). 
In addition, TRTM maintains efficient 
execution, achieving a lower mean execution time of 
0.82 minutes per version compared to 1.04 minutes for CTM.
}}

\section{\uppercase{Threats to Validity}}

\textbf{Construct Validity.} 
Construct validity concerns whether the evaluation metrics capture the intended notion of effectiveness. We use Accuracy and FDR, both widely adopted in prior TSM research~\cite{pan2023atm,pan2024ltm,siam2025exploratory}, where Accuracy measures the proportion of fault-revealing test cases preserved and FDR captures whether at least one such test case is retained.

Our approach relies on change history-based indicators (change frequency and change extent) to estimate class-level risk. These indicators have been extensively used in software evolution and defect prediction studies as proxies for fault-proneness \cite{arvanitou2017method,graves2002predicting}. By combining these indicators with temporal modeling and evaluating 
across multiple configurations and budgets, the assessment captures 
meaningful variations in effectiveness.

\noindent\textbf{Internal Validity.} 
Internal validity concerns whether the observed differences are attributable to the proposed approach rather than experimental artifacts. To ensure a fair comparison, all approaches were evaluated under identical settings, utilizing the same datasets, dependency extraction pipeline, and minimization budgets. While our analysis relies on static test--class dependency extraction—which may miss dynamic behaviors such as reflection or runtime polymorphism—this limitation applies uniformly across all evaluated configurations, thereby preserving the validity of the comparative results. Furthermore, TRTM is fully deterministic and does not rely on stochastic optimization components; this architectural choice, combined with a uniform execution environment, guarantees consistent and reproducible results across all experimental runs.

\textbf{Conclusion Validity.} 
Conclusion validity concerns the statistical soundness of the reported results. To support reliable conclusions, we employ non-parametric statistical tests: the Wilcoxon signed-rank test for paired comparisons of Accuracy, and Fisher’s exact test for fault detection outcomes (FDR). Effect sizes are reported using Cliff’s $\delta$ for Accuracy and odds ratios for FDR to quantify the magnitude of observed differences. These measures help ensure that the reported improvements are statistically supported and not due to random variation.

\textbf{External Validity.} 
External validity concerns the generalizability of the findings beyond the studied context. The evaluation is conducted on 14 Java projects from the Defects4J benchmark, comprising 631 buggy versions. While these projects are open-source and may differ from industrial systems, Defects4J is one of the most widely used benchmarks in software testing research due to its collection of real faults and reproducible test suites \cite{rjust2014defects4j}, supporting the relevance of our findings to the broader community.

Defects4J versions typically contain a single fault per buggy version, whereas real-world systems may involve multiple interacting faults. While this is a limitation, it provides a controlled setting for evaluating fault detection capability and has been widely adopted in prior TSM studies.

Additionally, buggy versions in Defects4J are often closely associated with recent changes, reflecting the bursty nature of software development and the temporal proximity between code modifications and fault introduction \cite{graves2002predicting,rahman2013how}. This aligns with established software evolution principles, where recently modified components are more likely to be fault-prone. A single global temporal horizon is adopted to enable consistent large-scale analysis, although the observed variation across projects suggests that temporal sensitivity may be project-dependent. Future work can explore adaptive, project-specific horizon selection.

Finally, although the study focuses on Java systems, the underlying 
principles of change history analysis and temporal risk modeling are 
not language-specific and can be extended to other ecosystems with 
appropriate tooling, and future work should evaluate the approach on 
more diverse datasets, including systems with different development 
patterns and multi-fault scenarios.

\section{\uppercase{Conclusions}}
\label{sec:conclusion}

In this study, we addressed a key limitation in change history-based black-box test suite minimization: the uniform, time-agnostic treatment of historical modifications. We introduced Temporal Risk-driven Test Suite Minimization (TRTM), the first approach to incorporate temporal modeling into change history-based TSM to capture the recency of software changes when estimating class-level risk. TRTM utilizes two change history-based indicators, change frequency and change extent, and applies temporal attenuation over configurable time horizons to compute time-weighted class-level risks, which are then aggregated through statistical operators to guide test suite minimization.

We evaluated TRTM across $2 \times 10 \times 4 = 80$ configurations, combining two change metrics, ten temporal horizons (1--512 days), and four aggregation operators, on a large dataset consisting of 14 Java projects with 631 buggy versions from Defects4J. The results show that TRTM consistently improves fault detection effectiveness across projects and minimization budgets. At the 50\% minimization budget, TRTM achieves a mean Accuracy of 0.72 (vs.\ 0.66) and FDR of 0.75 (vs.\ 0.69) compared to the state-of-the-art baseline CTM, with statistically significant improvements while maintaining efficient execution time.

Future work can investigate adaptive, project-specific temporal configurations to further optimize minimization effectiveness. Additionally, while the underlying principles of temporal risk modeling are inherently language-agnostic, our current evaluation focused exclusively on Java systems; therefore, future research should explore the applicability of TRTM across other programming languages. Finally, extending this evaluation to encompass multi-fault datasets and real-world industrial systems remains a crucial next step for establishing broader generalizability.

\bibliographystyle{apalike}
{\small
\bibliography{example}}
\end{document}